**Holographic principle tells why maximum entropy principle, lognormal distribution, self-preserving and self-organization appearing in system evolutions across science**

**Lina Wang**[*]

[1] Shanghai Key Laboratory of Atmospheric Particle Pollution and Prevention (LAP[3]), Department of Environmental Science and Engineering, Fudan University, Shanghai 200433, China

*Email:* wanglina@fudan.edu.cn

**Abstract**: By applying the holographic principle that comes from black hole physics and information entropy ($S_{ient}$) introduced by Claude Shannon in 1948, we derive the mechanism of how similar patterns or tendencies observed from evolution processes of different systems are formed, such as maximum entropy principle, power law and lognormal distributions, self-preserving and self-organization. Taking a system of atmospheric particles as an example, we prove both self-preserving of the geometric standard deviation (GSD) and lognormal distribution appearing on the system patterns is caused by maximum information entropy principle (MIEP); By deducing thermodynamic entropy ($S_{tent}$) is a form of $S_{ient}$, and building up a particle system approaching a part of the holographic screen, we found the information probability statistics of particle number size distribution are realized on the holographic screen, resulting in the presence of information entropic force ($F_{ient}$). $F_{ient}$ drives the system to evolve in the direction of MIEP. Owing to the initial state, the evolution can undergo an increasing $S_{ient}$ and enhanced discretization, or a decreasing $S_{ient}$ and weakened discretization. The latter process corresponds to the system evolution tendency of self-organization; by introducing $F_{ient}$ into the flow field, we construct the unification field for fluid system evolution, including $F_{ient}$, $S_{ient}$, the Newton potential ($\varnothing$), the diffusion process, and Navier–Stokes equations, which are important for airborne particle motion, indicating that the atmospheric particle dynamics could be driven by $F_{ient}$. By numerical simulations of particle collisions and coagulation based on the classic theory of integral–differential Smoluchowski equation, we prove the above conclusions that the evolution of the particle system is indeed information distributing, inherited from the initial state, measured by information probability, evaluated by $S_{ient}$, and driven by $F_{ient}$. We further prove for quantum mechanism, wave function also satisfies MIEP. The findings of this study indicate that based on holographic principle, it may be possible to guide the evolution of a given system towards certain desirable patterns and outcomes across many scientific disciplines.

## *1. Background.*

Similar patterns or tendencies are observed from various systems evolution, such as lognormal distributions, power law distribution, self-preserving and self-organizing. Lognormal distributions are continuous probability distributions of a random variable, the logarithm of which displays a normal distribution. As a statistical methodology, it has been successfully applied in representing system states across sciences [1]. From the perspective of statistics, self-preserving distribution is evaluated using the standard deviation, which represents the discretization degree of a system. It can be the result of either strengthening or weakening of discretization, as the final state has nothing to do with the initial state [2]. Namely, self-preserving can be the result of individual components developing into order or disorder, while self-organization (self-assembly) occurs when a spontaneous increase in order is produced from the cumulative interactions of individual components of disordered systems [3]. These concepts have been observed and proven useful for displaying states in diverse systems. However, each system is highly dynamic with individual components proceeding in random motions, it is challenging to find a universal mechanism driving evolutions of so many disciplines exhibiting identical patterns or tendencies. Limpert summarized 61 categories of lognormal distributions from disciplines including geology and mining, human medicine, environment, atmospheric sciences, and aerobiology, phytomedicine and microbiology, plant physiology, ecology, food technology, linguistics, social sciences, and economics He came up a question: what the underlying principles of permeability that cause lognormal variability?[4]

A framework for maximum information entropy principle (MIEP) proposed by Shannon and supplemented by Jaynes [5-7] sheds light on exploring probability distributions for individual components in a dynamic system [8-14]. For example, tree-like networks defined by the absence of loops, are characteristic of tributary river networks and found abundantly in nature across different systems and scales (e.g., botanical trees, veins of leaves, blood vessels, lightning, and river networks). Tejedora et al. (2017) proved tree-like networks can be interpreted using MIEP[15]. The basic idea is that the shape of the least biased distribution consistent with knowledge of the prior constraints of a given system can be inferred by MIEP. The recent growth of the number of publications related to the MIEP and the obtained interesting results in different areas of science, from physics to biology, are the best proof that the MEPP is scientific and is important.

In addition to MIEP, Martyushev reviewed the main areas of research that laid the foundations for the maximum entropy production principle (MEPP)[16]. This principle implies that the value of entropy production not only remains positive but is also maximized during the evolution of a nonequilibrium system. The review discussed its applications and limitations in the areas of theoretical physics, geophysics, and hydrodynamics, nonequilibrium crystallization, biology, material science, astrophysics, brain science and neuroscience. The review concluded that the MEPP is a universal and very fruitful approach in many branches of science, and suggests us to view the world from a common standpoint without dividing it into the living and nonliving. At the deepest levels of this world, some 'primitive' physicochemical processes occur, which obey the MEPP.

Although there are still controversial opinions on the link between MIEP and MEPP, the reason why MIEP produces the least biased predictions of distribution shapes and why MEPP can be found from different areas of science, and under what circumstances they can be applied, have not been elucidated.

Considering above gaps in knowledge: *i*) this study proves that lognormal distributions and power law distributions are indeed the distribution arising from MIEP for individual components in systems, and *ii*) self-preserving is a direct reflection of the state of MIEP. *iii*) Given that maximum entropy-driven phenomena or maximum entropy-stabilized reactions are reported across a variety of disciplines including biology, chemistry and materials [17-26] [27-30], and self-organizing systems can be found practically everywhere, some studies have employed the concepts of entropy and information[31] in order to understand them, we wonder whether there is an existing force driving diverse systems, to evolve along the same direction of maximum entropy via interactions among individual components to induce statistical regularities and reconstruct information distributions. The strongest supporting evidence for the holographic principle comes from black hole physics [32, 33]. So next, based on thermodynamic entropy being as a form of information entropy and building up a holographic screen, we derive a potential information entropy force ($F_{ient}$) driving the system evolution either by an increasing information entropy ($S_{ient}$) and enhanced discretization, or by a decreasing information entropy and weakened discretization. And *iv*) found the latter process corresponds to self-organization. *v*) Then we constructed the unification field for fluid system evolution by introducing $F_{ient}$. *vi*) And further proved above findings by numerical simulations of particle collisions and coagulation based on the classic theory of integral–differential Smoluchowski equation. *vii*) We also prove MIEP is applicable to the solution of Schrödinger equation in quantum mechanics. viii) We proposed three equations to describe the evolution state of any system. And the relevant promising directions were discussed. Overall, we found based on the holographic principle, the system evolution is indeed information distributing, inherited from the its initial state, measured by information probability, evaluated by $S_{ient}$, and driven by $F_{ient}$.

## 2. *Lognormal distribution and power law distribution obey MIEP.*

The information entropy ($S_{ient}$) for a given system with random variable $x$ {$x_1, x_2...x_n$} ($n \geq 1$) is shown as equation (2-1):

$$S_{ient} = -r \sum_{i=1}^{n} p_i * \ln(p_i) \qquad (2-1)$$

And the constraints are shown as equation (2-2):

$$\sum_{i=1}^{n} p_i = 1 \qquad (2-2)$$

where $r$ equals $\frac{1}{ln2}$, the probability of each variable is $p$ {$p_1, p_2,...p_n$}, ($0 \leq p_i \leq 1$, $i=1, 2,..., n$). The typical mathematical expression of the characteristics, $f_k$, namely, the probability of normalized constraints is shown as equation (2-3):

$$f_k = \sum_{i=1}^{n} p_i * f_i^{\ k} \ (k = 1,2,3,4...,m) \qquad (2-3)$$

where $k$ is the serial number for the typical characteristics, $k$ = 1, 2, 3……m; and $f_i^{\ k}$ is the $k^{th}$ typical characteristic of the $i^{th}$ subsystem.

For an airborne particle system, its size distribution ranges over several orders of magnitude, hence logarithmic coordinate ($lndp_i$) is always adopted to characterize the particle size distribution[34]. $p_i$ is the probability of particles belonging to the $i^{th}$ size bin, $f_i^k$ is equal to $(\ln(dp))^k$, and $f_k$ is the mean of $(\ln(dp))^k$. Lagrange multiplier methodology is always adopted to obtain solutions for extreme problems with constraint conditions. Hence the Lagrange function for obtaining MIEP of equation (2-1) is built up via the left side of the $k^{th}$ formula in equation (2-2) is multiplied by $\lambda_k - 1$, and the right side of equation (3) is multiplied by $\lambda_k$, as given by equation (2-4):

$$L(\lambda_k, \lambda, p_i) = -\sum_{i=1}^{n} p_i * \ln(p_i) - \sum_{k=1}^{m} \lambda_k \sum_{i=1}^{n} p_i * f_i^{(k)} - (\lambda - 1) \sum_{i=1}^{n} p_i \quad (2-4)$$

The partial derivative of equation (2-4) is shown as equation (2-5):

$$\frac{\partial L}{\partial p_i} = -1 - \ln p_i - \sum_{k=1}^{m} \lambda_k f_i^{(k)} - (\lambda - 1) \quad (2-5)$$

Setting equation (2-5) equal to 0 and replace $f_i^k$ with $(\ln(dp))^k$, the MIEP equation for airborne particle systems is obtained, as given by (2-6):

$$p_i = \exp(-\sum_{k=0}^{m} \lambda_k (\ln dp_i)^k) \quad (2-6)$$

$(\ln dp_i)^k$ is the $k^{th}$ origin moment, $k = 1,2...., m$; $\lambda_k$ is the Lagrange multiplier and can be obtained by the regress function. Accordingly, the information entropy for a given system with random variable $x$ $\{x_1, x_2,...x_n\}$, $(n \geq 1)$ can be determined from its typical characteristics, as shown in Table 1.

**Table 1 parameters for k$^{th}$ moments for particle number size distribution**

| m | Model | Parameters |
|---|-------|------------|
| 1 | $p_i = \exp(-\lambda_0 - \lambda_1 lndp_i)$ | $\lambda_0, \lambda_1$ |
| 2 | $p_i = \exp(-\lambda_0 - \lambda_1 lndp_i - \lambda_2(lndp_i)^2)$ | $\lambda_0, \lambda_1, \lambda_2$ |
| 3 | $p_i = \exp(-\lambda_0 - \lambda_1 lndp_i - \lambda_2(lndp_i)^2 - \lambda_3(lndp_i)^3)$ | $\lambda_0, \lambda_1, \lambda_2, \lambda_3$ |
| 4 | $p_i = \exp(-\lambda_0 - \lambda_1 lndp_i - \lambda_2(lndp_i)^2 - \lambda_3(lndp_i)^3 - \lambda_4(lndp_i)^4)$ | $\lambda_0, \lambda_1, \lambda_2, \lambda_3, \lambda_4$ |
| 5 | $p_i = \exp(-\lambda_0 - \lambda_1(lndp_i) - \lambda_2(lndp_i)^2 - \lambda_3(lndp_i)^3 - \lambda_4(lndp_i)^4 - \lambda_5(lndp_i)^5)$ | $\lambda_0, \lambda_1, \lambda_2, \lambda_3, \lambda_4, \lambda_5$ |
| … | … | … |

It is noted when taking *x* as a characteristic quantity instead of $lndp_i$ in table 1, the power law and normal distributions are realized by the first (*m* = 1) and second moment (*m* = 2) of MIEP, respectively. That is, the common patterns including normal, lognormal, and power law distributions arising across broad disciplines, are indeed subsets of MIEP distributions. When *m* takes on smaller values, the number of constraint equations is less, and the process of finding a solution is simple, and vice versa for *m* with greater values. As power law distributions and lognormal distributions are obtained widely from different scientific areas, MIEP is not only a mathematical procedure, it is also approved by amount of experiments.

## 3. Self-preserving obeys MIEP for lognormal and power law distributions.

*Lognormal distributions.*

The geometric standard deviation (GSD, $\sigma_g$) is a parameter for lognormal distributions, as given in equation (3-1):

$$f(lnx) = \frac{1}{\sqrt{2\pi}ln\sigma_g} e^{-\frac{(lnx-ln\mu)^2}{2*(Ln\sigma_g)^2}} \qquad (3-1)$$

We derive the relationship between the MIEP and $\sigma_g$ for the lognormal distribution based on equation (2-1), as shown by equation (3-2):

$$S_{ient} = -r \int p(x) \times lnp(x)\, dx \qquad (3-2)$$

For an airborne particle system with a lognormal probability density function, equation (3-3) is obtained:

$$p(lndp_i) = \frac{1}{\sqrt{2\pi}ln\sigma_g} \times e^{-\frac{(lndp_i-ln\mu)^2}{2\times(ln\sigma_g)^2}} \qquad (3-3)$$

Equation (3-3) is substituted into equation (3-2) to yield equation (3-4):

$$S_{ient} = r \int \frac{1}{\sqrt{2\pi}ln\sigma_g} \times e^{-\frac{(lndp_i-ln\mu)^2}{2\times(ln\sigma_g)^2}}$$
$$\times \left\{\frac{(lndp_i - ln\mu)^2}{2 \times (ln\sigma_g)^2} + ln(\sqrt{2\pi}ln\sigma_g)\right\} d(lndp_i) \qquad (3-4)$$

According to the definition of variance, equation (3-5) is obtained:

$$\int \frac{1}{\sqrt{2\pi}Ln\sigma_g} \times e^{-\frac{(Lndp_i-Ln\mu)^2}{2\times(Ln\sigma_g)^2}} \times \{(lndp_i - ln\mu)^2\}d(lndp_i) = (ln\sigma_g)^2 \qquad (3-5)$$

Therefore, equation (3-6) is derived:

$$S_{ient} = \frac{1}{2}r + r\int p(lndp_i) \times ln(\sqrt{2\pi}ln\sigma_g)\, d(lndp_i)$$
$$= \frac{1}{2}r + rl\,n(\sqrt{2\pi} \times ln\sigma_g) \qquad (3-6)$$

As r equals to $\frac{1}{ln2}$, equation (3-6) indicates that $\sigma_g$ of a system could remain constant when the system attains the steady state; that is, achieving its MIEP. Hence, $\sigma_g$ is an intrinsic property of a specific system, being limited by environmental constraints but not affected by initial conditions. Identification of $\sigma_g$ provides valuable insights for guiding the system development, and is applicable to a number of scientific disciplines. When the information of a steady state system is destroyed, its $S_{ient}$ evolves in the direction of entropy changing, resulting in self-preserving distribution disappearing, so that $\sigma_g$ starts to vary. When the constraints are set for a specific system, $\sigma_g$ is quantified, the mobility of all characteristic quantities can be assessed. That is, the system structure and variance components are obtained.

*Power law distribution.*

When taking *x* as a characteristic quantity instead of *ln(x)*, the power law is realized with the first (*m*=1) moment of MIEP. For a power law distribution, the deduction steps are equations (3-7)-(3-13). The exponential function is represented by equation (3-7):

$$y = \lambda e^{-\lambda x} \qquad (3-7)$$

According to the definition of information entropy, the information entropy of an exponential function is described according to equation (S2), and is transformed into equations (S3)-(S7):

$$S_{ient} = r\int_0^{+\infty} \lambda e^{-\lambda x} ln(\lambda e^{-\lambda x})\, dx \qquad (3-8)$$

$$S_{ient} = r \int_0^{+\infty} \lambda e^{-\lambda x}(ln\lambda - \lambda x)dx \qquad (3-9)$$

$$S_{ient} = r \int_0^{+\infty} \lambda \, ln\lambda \, e^{-\lambda x}dx - r \int_0^{+\infty} \lambda^2 x \, e^{-\lambda x}dx \qquad (3-10)$$

$$S_{ient} = -rln\lambda e^{-\lambda x}|_0^{+\infty} + r\lambda x e^{-\lambda x}|_0^{+\infty} + r \int_0^{+\infty} \lambda e^{-\lambda x}dx \qquad (3-11)$$

$$S_{ient} = rln\lambda + 0 - 0 - re^{-\lambda x}|_0^{+\infty} \qquad (3-12)$$

$$S_{ient} = rln\lambda + r \qquad (3-13)$$

Equation (3-13) is for all $S_{ient}$, such that it is also applicable to $S_{ient_{max}}$, as given by equation (3-14):

$$S_{ient_{max}} = rln\lambda + r \qquad (3-14)$$

According to the property of exponential distribution, the relationship between its variance and $\lambda$ is given as equation (3-15):

$$D(X) = \frac{1}{\lambda^2} = \sigma^2 \qquad (3-15)$$

From equations (3-13) and (3-15), equation (3-16) is obtained, indicating $\sigma$ takes on a certain value when the particle system reaches its maximum $S_{ient}$.

$$S_{ient_{max}} = -\frac{1}{2}rln(\sigma^2) + r \qquad (3-16)$$

Above all, self-preserving obeys MIEP for lognormal and power law distributions. Inspired by these findings that MIEP causes common patterns and tendencies to arise in different system states, we speculate that MIEP assumes the role of a force direction.

## 4. Quantification of information entropic force ($F_{ient}$).

Therefore, we assume the information associated with a part of space obeys the holographic principle [35, 36]. Based on the holographic principle and the equipartition rule, Verlinde [37] proposed a thermodynamic entropic force. Being inspired by his methodology, we quantify a more general entropic force, $F_{ient}$. Firstly, the relationship between $S_{ient}$ and thermodynamic entropy ($S_{tent}$) is given as equation (4-1):

$$S_{ient} = -r \sum_{i=1}^{n} p_i * ln(p_i) = -r \sum_{i=1}^{n} \frac{1}{\Omega} * ln\left(\frac{1}{\Omega}\right) = r\, ln\Omega \rightarrow S_{tent} = k_B ln\Omega \quad (4-1)$$

where $k_B$ is Boltzman's constant. The equal probability hypothesis of the equilibrium state for an isolated system is considered as $p_i = 1/\Omega$, and $\Omega$ is the total number of microscopic states. Hence, equation (4-2) is obtained:

$$S_{ient} = \frac{r}{k_B} \cdot S_{tent} \quad (4-2)$$

Equation (4-2) shows $S_{tent}$ can be obtained from $S_{ient}$, but irreversible, indicating $S_{ient}$ is a more inclusive entropy. Namely, the principles of $S_{tent}$ can be applied to $S_{ient}$ when considering thermodynamic information. We consider $S_{ient}$ includes $S_{tent}$ and the nonthermodynamic entropy($S_{ntent}$), which represents the information entropy of thermodynamic and nonthermodynamic systems, respectively.

Next, a small piece of a holographic screen is constructed in the information field. The change of information amount (*dD*) at a point near to the screen can be obtained by the information entropy change ($dS_{ient}$) and the information potential ($\emptyset_{ient}$) at that point, as given by equation (4-3):

$$dD = \emptyset_{ient} \cdot dS_{ient} \qquad (4\text{-}3)$$

For thermodynamic systems, $dD$ is heat transfer, $\emptyset_{ient}$ is thermodynamic temperature, and $dS_{ient}$ is thermodynamic entropy change ($dS_{tent}$). While in the nonthermodynamic field, $F_{ient}$ describes the change of information amount caused by the interactions among information, as given by equation (4-4):

$$F_{ient} \cdot dx = \emptyset_{ient} \cdot dS_{ient} \qquad (4\text{-}4)$$

Similarly, for a system of particles with different sizes, we assume the particle system with total mass of *m* approaching a holographic screen, as shown in Figure 1.

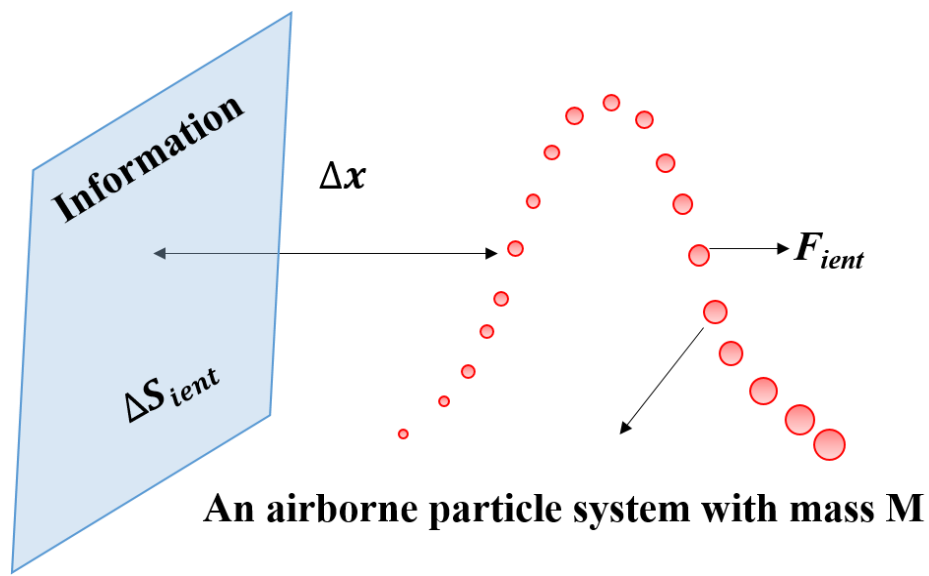

**An airborne particle system with mass M**

**Figure 1 A system of particles with different sizes approaching a part of the holographic screen**

The particles receive the change of information on the screen [32]. Motivated by Verlinde's argument based on Bekenstein, equation (4-5) is obtained:

$$\frac{k_B}{r}\Delta S_{ient} = 2\pi k_B \frac{mc}{h}\Delta x \quad (4-5)$$

where $m$ is the mass of the particles, $c$ is the speed of light, and $h$ is the Planck constant. $S_{ient}$ of a system depends on the distance $\Delta x$ to the screen, $F_{ient}$ could arise in an analogous biophysical system, as given by equation (4-6):

$$T\frac{k_B}{r}\Delta S_{ient} = F_{ient}\Delta x \quad (4-6)$$

where $T$ is the temperature of the holographic screen. Equation (4-6) indicates when the information of a system is realized on the holographic screen, $F_{ient}$ causes information variation, resulting in alteration of the relative distance $\Delta x$ to the screen, followed by energy exchange, temperature change, and $S_{ient}$ variation. According to equation (4-5) and (4-6), $F_{ient}$ can be given as equation (4-7):

$$F_{ient} = 2 \times \pi \times m \times T \times \frac{k_B c}{h} \quad (4-7)$$

Next, we assume $E$ is distributed on a spherically shaped holographic screen with radius $R$, and mass $M$ is located at the origin of a coordinate system as the source mass. Equipartition rule is introduced to define temperature $T$ [38], the equality of energy and mass, and the holographic principle to give the number of states $N$, as equations (4-8)-(4-10):

$$E = \frac{1}{2}Nk_B T \quad (4-8)$$

$$E = Mc^2 \quad (4-9)$$

$$N = \frac{Ac^3}{Gh} \tag{4-10}$$

The area of the holographic screen is $A = 4\pi R^2$. $T$ can be determined from equations (4-8)-(4-10), as shown in equation (4-11):

$$T = \frac{GMh}{2\pi k_B c R^2} \tag{4-11}$$

Substituting (4-11) into (4-7), $F_{ient}$ is realized as Newton's law of gravity:

$$F_{ient} = \frac{GmM}{R^2} \tag{4-12}$$

The relationship between acceleration *a* and temperature has been shown by Unruh (36), as given by equation (4-13):

$$T_U = \frac{a}{2\pi k_B} \frac{h}{c} \to T \tag{4-13}$$

where $T_U$ is the bulk temperature, and $T$ is the boundary surface temperature. Hence, equations (4-7) and (4-13) lead to Newton's second law, as given by equation (4-14):

$$F_{ient} = ma \tag{4-14}$$

Equations (4-12) and (4-14) indicate that gravitational attraction could be the result of the information of material objects being organized in space. $F_{ient}$ causes variations of the statistical behaviors associated with versatile degrees of information freedom for different systems encoded on the holographic screen, leading all systems to develop in the same direction of MIEP.

We also derived the relationship between $S_{ient}$ and *a*. Assume a particle with mass *m* near to a holographic screen. Each bit of the screen carries energy of $\frac{1}{2} k_B T$. The total number of bits on the screen is *n* and follow equation (4-15):

$$mc^2 = \frac{1}{2} n k_B T \tag{4-15}$$

From equations (4-5), (4-11), and (4-15), equation (4-16) is obtained:

$$\frac{\Delta S_{ient}}{nr} = \frac{a\Delta x}{2c^2} \tag{4-16}$$

Equation (4-16) indicates that the acceleration, *a*, and the entropy gradient, $\frac{\Delta S}{\Delta x}$, are closely related. When *a*=0, $S_{ient}$ reaches up to its greatest value, so that the system remains constant. Inertia is a consequence of the fact that a particle in rest will stay in rest because there is no entropy gradient. The relationship between *a* and ∅ can be given as equation (4-17):

$$a = -\nabla\emptyset \tag{4-17}$$

Therefore, the relationship between $S_{ient}$ and ∅ can be given as equation (4-18):

$$\frac{\Delta S_{ient}}{nr} = -\frac{\Delta\emptyset}{2c^2} \tag{4-18}$$

Equation (4-18) shows that ∅ keeps track of the depletion of entropy per bit. The holographic direction is given by the gradient ∆∅. In other words, the holographic screen corresponds to an equipotential surface.

Next, Poisson equation is adopted to describe a general matter distribution. A holographic screen is chosen corresponding to an equipotential surface with fixed $\emptyset_0$ and a static matter density of $\rho(\vec{t})$. Equation (4-19) can be obtained from equations (4-13) and (4-17):

$$k_B T = \frac{1}{2\pi} \frac{h\nabla\emptyset}{kc} \tag{4-19}$$

where *k* is a fixed coefficient. Equation (4-10) can be generalized into equation (4-20):

$$dN = \frac{c^3}{Gh} dA \qquad (4\text{-}20)$$

Equation (4-8) can be given in its integral form, as shown by equation (4-21):

$$E = \frac{1}{2} k_B \int T dN \qquad (4\text{-}21)$$

Energy $E$ is again expressed in terms of the total enclosed mass $M$, and Gauss's law is adopted to obtain equation (4-22):

$$M = \frac{1}{4\pi G} \int \nabla \emptyset dA \qquad (4\text{-}22)$$

The Poisson equation is given as equation (4-23):

$$\nabla^2 \emptyset(\vec{t}) = 4\pi G \rho(\vec{t}) \qquad (4\text{-}23)$$

Then we use the relationship between the arbitrary infinitesimal displacements $\delta \vec{t_i}$ of particles and the resulting entropy change to obtain the force acting on the matter particles that are located at arbitrary points outside the screen. The change of entropy density can be expressed as equation (4-24):

$$\delta s = k_B \frac{\delta \emptyset}{2c^2} dN \qquad (4\text{-}24)$$

where $\delta \emptyset$ is the response of $\emptyset$ due to the shifts, $\delta \vec{t_i}$, of the particle positions, as shown in Figure 2.

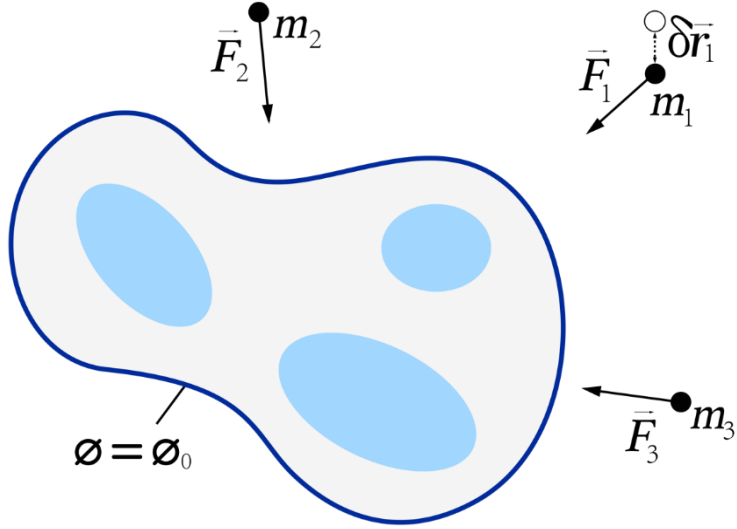

**Figure 2 Particles are located at arbitrary points and influenced by $F_{ient}$.**

$\delta\emptyset$ is calculated by solving the variation of the Poisson equation, as given by equation (4-25):

$$\nabla^2 \delta\emptyset(\vec{t}) = 4\pi G \sum_i m_i \vec{\delta t_1} \cdot \nabla_i \delta(\vec{t} - \vec{t_1}) \qquad (4-25)$$

$F_{ient}$ on the particle system can be obtained using the local temperature and the variation of $S_{ient}$, as given by equation (4-26):

$$\sum_i \overrightarrow{F_{ient_i}} \cdot \vec{\delta r_i} = \int T \delta S_{ient} \qquad (4-26)$$

The variation of $\emptyset$ can be obtained from Green's function for the Laplacian. The basic identity that needs to be proved is demonstrated by equation (4-27):

$$\sum_i \overrightarrow{F_{ient_i}} \cdot \vec{\delta r_i} = \frac{1}{4\pi G} \int (\delta\emptyset \nabla\emptyset - \emptyset \nabla\delta\emptyset) dA \qquad (4-27)$$

Equation (4-27) holds for any location on the screen outside of the mass distribution, which indicates by realizing the information probability distribution of a system on a holographic screen, the system state can be described using $S_{ient}$ and $\emptyset$. These two parameters are determined by $F_{ient}$. Evolution of a system is driven via reconstructing information probabilities of the individual components that comprise the system. At this point, by applying the holographic principle, we could provide an interpretation for the large number of maximum entropy-driven phenomena or maximum entropy-stabilized reactions observed in various disciplines [17-26].

## 5. Unification of $F_{ient}$, $S_{ient}$, and $\emptyset$ in the flow field.

Considering diffusion is the major dynamic process of airborne particles, and the motion is associated with flow field. Also motived by the work of Leonard Susskind [35], in which he shows that the world can be represented as a hologram, we intend to apply $F_{ient}$, $S_{ient}$, and $\emptyset$ into flow fields. Firstly, particle diffusion equation is shown in equation (5-1):

$$J = \vec{i} \cdot J_x + \vec{j} \cdot J_y + \vec{k} \cdot J_z = -D\left(\vec{i} \cdot \frac{\partial n}{\partial x} + \vec{j} \cdot \frac{\partial n}{\partial y} + \vec{k} \cdot \frac{\partial n}{\partial z}\right) \quad (5-1)$$

where $J$ is the diffusion flux, $kg/(m^2 \cdot s)$, $n$ is the particle number concentration at a specific point, $\#/m^3$, $D$ is the diffusion coefficient. For one-dimensional diffusion, $J$ can be given as equation (5-2):

$$J_x = n \cdot v_x \quad (5-2)$$

where $v_x$ is the velocity in $x$ direction. Substituting equation (5-2) into equation (5-1) and taking the time derivative, equation (5-3) is obtained:

$$\vec{i} \cdot \left(v_x \frac{\partial n}{\partial t} + na_x\right) + \vec{j} \cdot \left(v_y \frac{\partial n}{\partial t} + na_y\right) + \vec{k} \cdot \left(v_z \frac{\partial n}{\partial t} + na_z\right) = -\left(\vec{i} \cdot \frac{\partial^2 Dn}{\partial x \partial t} + \vec{j} \cdot \frac{\partial^2 Dn}{\partial y \partial t} + \vec{k} \cdot \frac{\partial^2 Dn}{\partial z \partial t}\right) \quad (5-3)$$

Equation (5-4) can be obtained from equations (4-17) and (5-3):

$$\vec{i} \cdot \left(v_x \frac{\partial n}{\partial t}\right) + \vec{j} \cdot \left(v_y \frac{\partial n}{\partial t}\right) + \vec{k} \cdot \left(v_z \frac{\partial n}{\partial t}\right) - n\nabla \emptyset = -\left(\vec{i} \cdot \frac{\partial^2 Dn}{\partial x \partial t} + \vec{j} \cdot \frac{\partial^2 Dn}{\partial y \partial t} + \vec{k} \cdot \frac{\partial^2 Dn}{\partial z \partial t}\right) \quad (5-4)$$

Secondly, by analogy with diffusion driven by concentration differences, we consider the impact of fluid pressure differences, and apply $F_{ient}$ into N-S equation. The vector form of the N-S equation and the Reynolds number are given as equations (5-5) and (5-6):

$$\rho \frac{dv}{dt} = -\nabla p + \rho F + \mu \Delta v \quad (5-5)$$

$$\mathrm{Re} = \frac{\rho v d}{\mu} \quad (5-6)$$

where $\rho$ is the fluid density, $v$ is the fluid velocity, $p$ is the fluid pressure, $\mu$ is the dynamic viscosity of the fluid, $d$ is the characteristic length of the environment in which the fluid is flowing, and $F$ is the external force acting on the fluid. Fluid flow is under frictionless resistance, so equation (5-5) is simplified into equation (5-7):

$$\rho a = -\nabla p + \rho F \quad (5-7)$$

We combine equations (4-17), (5-6), and (5-7), and then take the time derivative, equation (5-8) and (5-9) are obtained:

$$-\rho \nabla \emptyset = -\nabla p + \rho F \quad (5-8)$$

$$\frac{d\mathrm{Re}}{dt} = -\frac{\rho d \nabla \emptyset}{\mu} \quad (5-9)$$

According to equations (5-8), (5-9) ) and $\frac{\Delta S_{ient}}{nr} = -\frac{\Delta \emptyset}{2c^2}$, fluid flow follows the direction of increasing $S_{ient}$, driven by the resultant forces ($F_{ient}$) of fluid mass. The temporal variation rate of *Re* is positively correlated with the entropy gradient, but negatively correlated with the potential gradient. Above all, under the holographic theory framework, $S_{ient}$, the Newtonian potential field, the flow field, and $F_{ient}$ are unified.

## 6. $F_{ient}$ drives the evolution of airborne particle systems.

Analogous to the infiltration process, $F_{ient}$ drives collisions and diffusions of airborne particles during their evolution in the unification fluid field after being released from an emission source, as shown in Figure 3.

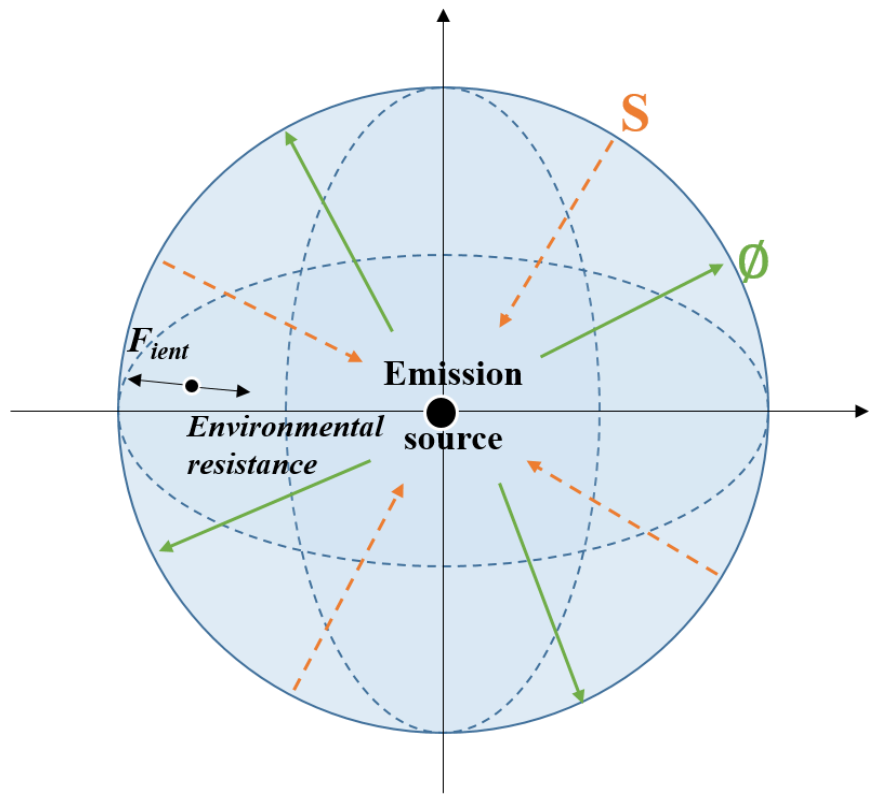

**Figure 3 An airborne particle released from the emission source undergoing collision and diffusion caused by $F_{ient}$**

The holographic screen for airborne particles is taken as an equipotential spherical surface, and the process of particles being released to the environment is entropy changing. $\emptyset$ decreases along the sphere radius, but $S_{ient}$ increases along the sphere radius. All particles affected by their respective resistances at the same time eventually evolve into a steady state, namely, MIEP, at which state a balance between the generated information entropy within the system ($\frac{d_i s}{dt}$) and the change of information entropy caused by the external information entropy impacting on the internal system ($\frac{d_e s}{dt}$) is achieved. The self-preserving phenomenon for the entire system occurs. During this process, airborne particles experience uneven collisions from surrounding molecules, so that coagulation occurs owning to a spherically symmetric process of collision and diffusion. Driven by the unbalanced collision forces, $F_{ient}$, the particle system evolves in the direction of changing entropy.

This section adopts the classical theory based on the concept of Newtonian force to calculate particle dynamic processes in a closed system with two different initial conditions for *GSD*. Initial conditions: $k_b$=1.38×10$^{-23}$ J/K, *T*=298.15K, $\mu$=1.83245×10$^{-5}$pa·s, initial concentration ρ=1000kg/m$^3$, the two scenarios of initial conditions have different *GSD*, $w_1 = \sqrt{ln\frac{4}{3}}$, $w_2 = \sqrt{ln\frac{10}{3}}$, $v_g = \frac{\sqrt{3}}{2}$, $K_b$ is the Boltzmann constant, *T* is the system thermodynamic temperature, *μ is the particle kinetic viscosity,* ρ is for particle density, *w$_1$* and *w$_2$* are dimensionless volume geometric standard deviation of particles for the first time simulation and the second time simulation, respectively, $v_g$ is the dimensionless particle average volume. Here we prove when the airborne particle system achieves the state of MIEP, its GSD remains constant, and during this process, the information entropy can be either increase or decrease, but the final state is certain for the closed system. The simulation is based on Smoluchowski's groundbreaking work for coagulation processes [39, 40] according to integral–differential SE. Particle dynamic processes is calculated as equation (6-1):

$$\frac{\partial n(v,t)}{\partial t} = \frac{1}{2}\int_0^v \beta(v_1, v-v_1)n(v_1,t)n(v-v_1,t)dv_1 - n(v,t)\int_0^\infty \beta(v_1, v)n(v_1,t)dv_1 \quad (6\text{-}1)$$

in which $n(v, t)\,dv$ is the number of particles whose volume is between $v$ and $v + dv$ at time $t$, and $\beta(v_1, v)$ is the collision kernel for two particles of volumes v and v'. The initial particle number size distribution obeys a lognormal distribution. Equation (38) cannot be solved, so it is multiplied by $v^k$, and by integrating $v$, and becomes into equation (6-2), which describes how each moment changes over time. This equation can be solved.

$$\frac{dm_k}{dt} = \frac{1}{2}\int_0^\infty \int_0^\infty [(v+v_1)^k - v^k - v_1^k]\beta(v,v_1)n(v,t)n(v_1,t)dvdv_1 \quad (6-2)$$

Each moment is defined as equation (6-3):

$$m_k = \int_0^\infty v^k n(v)dv \quad (6-3)$$

The Taylor-expansion method of moments (TEMOM) is introduced in solving Eq. (6-2) with the closure model for the $k^{th}$ moment (Yu et al., 2008), as given by equation (6-4):

$$m_k = \left(\frac{u^{k-2}k^2}{2} - \frac{u^{k-2}k}{2}\right)m_2 + (-u^{k-1}k^2 + 2u^{k-1}k)m_1 + \left(u^k + \frac{u^k k^2}{2} - \frac{3u^k k}{2}\right)m_0 \quad (6-4)$$

where u is the Taylor expansion point, defined to be $\frac{m_1}{m_0}$.

Therefore, in the **free molecule regime**, the collision kernel was derived from the kinetic theory of gases, as given by equation (6-5).

$$\beta(v, v_1) = B_1 \left(\frac{1}{v} + \frac{1}{v_1}\right)^{\frac{1}{2}} \left(v^{\frac{1}{3}} + v_1^{\frac{1}{3}}\right)^2 \quad (6-5)$$

where $B_1 = \left(\frac{3}{4\pi}\right)^{\frac{1}{6}} \left(\frac{6k_b T}{\rho}\right)^{\frac{1}{2}}$. Hence, according to equations (6-3)-(6-5), in the free molecule regime, equation (6-1) can be written as equations (6-6)-(6-8) [41]:

$$\frac{dm_0}{dt} = \frac{\sqrt{2}B_1\left(65m_2^2 m_0^{23/6} - 1210 m_2 m_1^2 m_0^{17/6} - 9223 m_1^4 m_0^{11/6}\right)}{5184 m_1^{23/6}} \quad (6-6)$$

$$\frac{dm_1}{dt} = 0 \quad (6-7)$$

$$\frac{dm_2}{dt} = -\frac{\sqrt{2}B_1\left(701 m_2^2 m_0^{11/6} - 4210 m_2 m_1^2 m_0^{5/6} - 6859 m_1^4 m_0^{-1/6}\right)}{2592 m_1^{11/6}} \quad (6-8)$$

Similarly, for the **Stokes region**, the collision kernel can be expressed as equation (6-9):

$$\beta_C = B_2 \left(\frac{1}{v^{1/3}} + \frac{1}{v_1^{1/3}}\right)\left(v^{1/3} + v_1^{1/3}\right) \quad (6-9)$$

where $B_2 = \frac{2k_b T}{\mu}$, and $\mu$ is the viscosity for ambient air. Equations (6-5) and (6-9) also indicate that the higher the temperatures are, the greater the collision frequencies will be, a prediction that is consistent with the conclusion of temperature being in direct proportion to $F_{ient}$. By inputting equation (6-9) into equation (6-2), equations (6-10)-(6-12) are obtained:

$$\frac{dm_0}{dt} = \frac{B_2(-151m_1^4 + 2m_2^2 m_0^2 - 13m_2 m_1^2 m_0)m_0^2}{81m_1^4} \quad (6-10)$$

$$\frac{dm_1}{dt} = 0 \quad (6-11)$$

$$\frac{dm_2}{dt} = -\frac{2B_2(-151m_1^4 + 2m_2^2 m_0^2 - 13m_2 m_1^2 m_0)}{81m_1^2} \quad (6-12)$$

For the entire size range, the methodology given by Otto et al. (1999) is applied [42], yielding equation (6-13):

$$\left(\frac{dm_k}{dt}\bigg| entire\right) = \left(\frac{dm_k}{dt}\bigg| co\right) \frac{1 + Kn_{m_k}}{1 + f(\sigma)Kn_{m_k} + 2Kn_{m_k}^2} \quad (6-13)$$

where $Kn_{n_{m_k}}$ and $f(\sigma_g)$ are given by equations (6-14) and (6-15)

$$Kn_{Mk} = \frac{1}{2}\left(\frac{dm_k}{dt}\bigg| co\right)\left(\frac{dm_k}{dt}\bigg| fm\right)^{-1} \quad (6-14)$$

$$f(\sigma_g) = 2 + 0.7\ln(\sigma_g)^2 + 0.85\ln(\sigma_g)^3 \quad (6-15)$$

Therefore, according to equations (6-6)-(6-15), the time-dependent particle number size distribution and corresponding *GSD* are obtained. Furthermore, based on the particle number size distribution, the time-dependent information entropy is given. The tendencies of *GSD* and information entropy are identical, as shown in Fig.4. *GSD1*, *S1*, and *GSD2*, *S2* are geometric standard deviation and information entropy for first time and second time simulation, respectively. Equation (6-16) is given by Yu et al. (2009) [43]:

$$(ln\sigma_g)^2 = \frac{1}{9}\left(\frac{m_0 m_1}{m_1^2}\right) \tag{6-16}$$

By inputting equation (6-16) into equation (3-6), the relationship between MIEP and the dynamic process is constructed, as shown in equation (6-17):

$$S_{ient} = \frac{1}{2}r + \frac{r\ln}{3}\left(\sqrt{2\pi\left(\frac{m_0 m_1}{m_1^2}\right)}\right) \tag{6-17}$$

The time-dependent of $S_{ient}$ and GSD calculated by simulated particle size spectrums are shown in Figure 4.

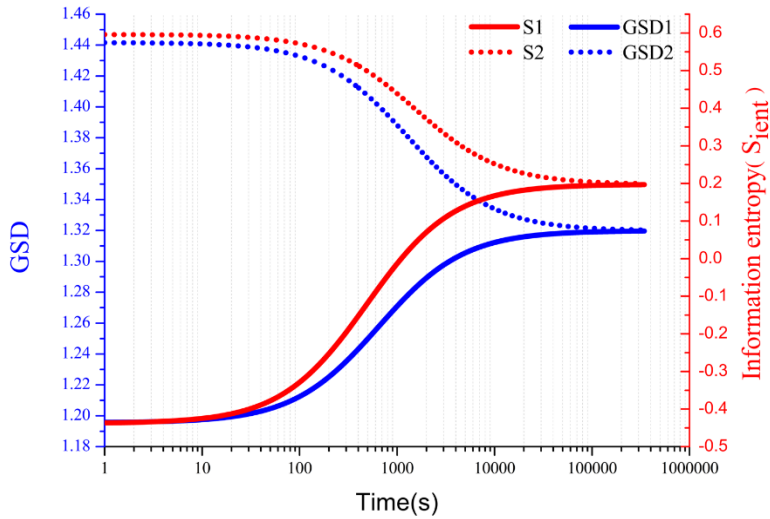

**Figure 4 Time-dependent variation of $S_{ient}$ based on the particle size spectrum, and time-dependent variation of GSD obtained by integral–differential SE**

Figure 4 provided two initial conditions with different initial GSD values. It is stressed that no matter what the initial conditions, the system with constant temperature will attain the same final state. This is consistent with previous discussions on $S_{ient}$, temperature, and GSD. During the evolution, GSD and $S_{ient}$ have identical variation tendencies, indicating that the airborne particle system is driven by $F_{ient}$, evolving in the direction of MIEP, and giving rise to a self-preserving size distribution. If the initial $S_{ient}$ is larger than the maximum $S_{ient}$ determined by the environmental constraints, $S_{ient}$, or to say GSD, shows a decreasing trend, which is affected by the external force of the system, as caused by gaseous molecules surrounding airborne particles, leading to the information entropy within the system flowing to the outside of the system. And vice versa for the scenario of the initial $S_{ient}$ is less compared to the maximum $S_{ient}$, in which both $S_{ient}$ and GSD increase until reach the steady state. This process is self-organization. GSD is not only a parameter that indicates the current state of system development, but also an apparent feature of $F_{ient}$.

## 7. Quantum mechanism

Quantum mechanics was one of the great discoveries for last century, which was proposed by Max Planck in 1900. In 1926, Erwin Schrödinger established the famous Schrödinger equation. By solving the equation, wave function is obtained, which has been used to describe the properties and motions of electrons in atoms and molecules. Therefore, various information of the system can be obtained from the wave function. However, the physical meaning of wave function is not clear so far, and wave function cannot be observed by experiment. It is difficult to solve the equation with the increase of electron number.

Maximum information entropy is one of the core concept of information theory, which was proposed by Shannon and supplemented by Jaynes [5-7]. Maximum information entropy explores the knowledge of probability distributions for individual components in a dynamic system [8-14]. The basic idea is that the shape of the least biased distribution consistent with knowledge of prior constraints of a given system can be inferred by maximum information entropy.

From the perspective of information theory itself, information entropy has some essential connection with the uncertainty relation in quantum mechanics, because information itself reflects the uncertainty of the system. Therefore, we prove wave equation is indeed the maximum information entropy distribution of Schrödinger equation.

*Wave equation obeys maximum information entropy.*

The most important equation of quantum mechanism is Schrodinger equation, as given by equation (7-1) for stationary condition:

$$-\frac{\hbar^2}{2m}\nabla^2 \times \psi(x) + V(x) \times \psi(x) = E \times \psi(x) \qquad (7\text{-}1)$$

For one-dimensional stationary state and infinite square well, solution is odd parity. Wave function is given as equation (7-2):

$$\psi(x) = \sqrt{\frac{2}{a}} \sin\frac{n\pi x}{a} \qquad (7\text{-}2)$$

The probability of Schrodinger equation is given as equation (7-3):

$$\text{pi}(x) = \psi(x)^2 = \frac{2}{a}\left(\sin\frac{n\pi x}{a}\right)^2 \qquad (7\text{-}3)$$

Only numerical solution is given from the point of mathematics only. There have been no mechanism interpretations. The solution is given as equation (7-4):

$$f(x) = \sum a_n x^n \qquad (7\text{-}4)$$

We notice the format of equation (7-4) is similar to equation (2-3). If we can prove ψ(logx) = ln (pi(x)), then we can prove equation (7-4) is indeed the maximum information entropy of stationary Schrodinger equation, namely, capable of providing an interpretation for quantum mechanism.

Based on equation (7-2) and (7-3), we obtain equation (7-5) and (7-6):

$$\psi(\log x) = \sqrt{\frac{2}{a}} \sin \frac{n\pi \log x}{a} \qquad (7\text{-}5)$$

$$\ln(pi(x)) = \ln \frac{2}{a} \left(\sin \frac{n\pi x}{a}\right)^2 \qquad (7\text{-}6)$$

We conduct series expansion near x = 0 for sin (x), and obtain equation (7-7):

$$\sin(x) = 0 + \frac{x}{1!} + \frac{x^{\wedge}3}{3!} + \frac{x^{\wedge}5}{5!} + \cdots \qquad (7\text{-}7)$$

Therefore, equation (7-5) is transformed into equation (7-8):

$$\psi(\log x) = \sqrt{\frac{2}{a}} [0 + \frac{1}{1!} * \frac{n\pi \log x}{a} + \frac{1}{3!} * \left[\frac{n\pi \log x}{a}\right]^3 + \frac{1}{5!} * \left[\frac{n\pi \log x}{a}\right]^5 + \cdots] \qquad (7\text{-}8)$$

As logx=$\frac{lnx}{ln10}$, then equation (7-8) is transformed into equation (7-9):

$$\psi(\log x) = \sqrt{\frac{2}{a}} [0 + \frac{1}{1!} * \left[\frac{n\pi}{a}\right] * \frac{lnx}{ln10} + \frac{1}{3!} * \left[\frac{n\pi}{a}\right]^3 * \left[\frac{lnx}{ln10}\right]^3 + \frac{1}{5!} \left[\frac{n\pi}{a}\right]^5 * \left[\frac{lnx}{ln10}\right]^5 + \cdots] \qquad (7\text{-}9)$$

From equation (7-6), we obtain equation (7-10):

$$\ln(pi(x)) = 2ln\sqrt{\frac{2}{a}} [0 + \left(\frac{n\pi}{a}\right) * x + \frac{1}{3!} * \left(\frac{n\pi}{a}\right)^3 * x^3 + \frac{1}{5!} * \left(\frac{n\pi}{a}\right)^5 * x^5 + \cdots] \qquad (7\text{-}10)$$

Therefore, we proved ψ(logx) = ln (pi(x)). That is, Quantum mechanism meets up MIEP. Based on the maximum information entropy, we prove the solutions of Schrödinger equation in quantum mechanics obeys the maximum entropy of information probability.

## 8. Discussions.

By considering MEPP is a form of MIEP and building up a holographic screen, we derive $F_{ient}$ either drives the increase of $S_{ient}$ and enhances discretization, or drives the decrease of $S_{ient}$ (self-organization) and weakens discretization in a system evolution, eventually reaching the state of MIEP associated with equilibrium between entropy generation inside system and entropy flow into outside system. This is because for a fixed system with the steady state, its MIEP is certain. At this point, power law and lognormal (normal) distributions, self-preserving and self-organization can be attained. To sum up, systems evolution is measured by information probability ($p_i$), evaluated by information entropy ($S_{ient}$) and potential ($\emptyset$), and ultimately caused by the information entropic force ($F_{ient}$), which is counteracted by environmental constraints. Take an airborne particle system as an example, environmental constraints include temperature, fluid viscosity, particle properties, etc. According to the Stokes resistance formula and the resistance formula of molecular thermal motion[44], the resistance scales are associated with the size of the particles. Here we define these parameters, which are capable of describing the evolution state of any system, as equation set (8-1):

$$\begin{cases} p_i = \exp(-\sum_{k=0}^{m} \lambda_k X(C)^{(k)}) \\ S_{ient} = -r \times \sum_{i=1}^{n} p_i Ln(p_i) \sim a \times \ln\Omega(C) \\ \overrightarrow{F_{ient}} \propto \nabla S \propto -\nabla\emptyset \end{cases} \quad (8-1)$$

$p_i$ is the information probability. $X(C)$ is the characteristic quantity expressed in its new format under constraints *C*, such as *lnX*. $\Omega$ relates to variables in the system. $a$ is $K_b$ in thermodynamic entropy, and the nonthermodynamic-factor constant in nonthermodynamic entropy, which is determined by interactions among objects under nonthermodynamic constraints, *C*. For the nonthermodynamic system, we introduce two definitions: nonthermodynamic development potential and acceleration. Only considering the impact of $F_{ient}$, nonthermodynamic development also satisfies $a=-\nabla\emptyset$, and is limited by nonthermodynamic environmental constraints. The findings of this study indicate that based on holographic principle, it may be possible to guide the evolution of a given system towards certain desirable patterns and outcomes across many scientific disciplines, which provides the possibility of studying phenomena at multiple scales under the same formalism. While the content of this study can be applied in the many natural sciences and social sciences, such as: galaxy, climate, rivers and mountains, economics, medicine, etc. Author would like to stress the following points:

*Big data and machine learning.*

Big data and machine learning has become ubiquitous and indispensable for solving complex problems in most sciences. Machine learning methods are particularly effective in situations where deep and predictive insights need to be uncovered from a large set of diverse data. There are number of studies found the prediction accuracy have been improved after adopting the principle of maximum entropy (MEPP and MIEP). However, the deep mechanism behind for why machine learning can predict future and why maximum entropy can improve the accuracy have not been provided. Readers cannot help to think MEPP and MIEP maybe just mathematical methodologies. While according to the previous sections, if the world can be represented as a hologram and the system evolution can be explained by the holographic principle, the information of all the system can be projected to a holographic plane. Then the $F_{ient}$ can drive the system evolve into the direction of maximum information entropy. If the dataset is large and diverse, the information of the system for the past can better describe the system itself and the $F_{ient}$ it provided is robust, which can predict the system revolution very well. It is noted the conclusion is based on the conditions that the constraints of the system are identical as previously. Otherwise, the new changes should be considered.

*Global temperature change and extreme weathers.*

Our earth is with billions of years old. Its evolution is very long time. Therefore, if we have recorded the earth data and climate data, their futures and evolutions can be predicted. Another point is based on our previous analysis that $S_{ient}$ is associated with temperature, and GSD of is an index for $S_{ient}$ in different systems. We can prove predict the global temperature changes according to the GSD of airborne particles in atmosphere. This is because that although the age of atmosphere is controversial, it maybe even earlier than the earth. Of course, other qualified systems related to surface temperature work as well.

*The Brain and consciousness science*

Brain science and neuroscience investigate the architecture of the brain and maps how each individual neuron operates. However, advances in brain science are relatively new. There is large amount of evidence that adopting entropy for quantification of the brain activity has yielded promising results: the (altered) state of consciousness, the ageing brain, and the quantification of the brain networks' information processing. MIEP can be a promising measure to study the complexities in brain science. The movie *The Wandering Earth II* was released during Chinese New Year in 2023. The character *Hengyu Tu* lost his wife and daughter in a car accident. He uploaded the brain data of his daughter successively to the computer of 550A and 550W, and gave his daughter a full digital life. Based on the findings of this study, his daughter is the real daughter who can think, speak, laugh, and cry, but cannot touch.

## 9. Conclusion.

Based on the conclusion of this study, we can rethink some of issues: Is the expanding universe the result of entropy increasing? Will the expanding stop? Is the evolution theory a result of the development of systems, owing to various constraints appear during the system development, so that is it indeed self-organization? This finding resonates with the question proposed by Schrödinger in What Is Life? The Physical

Aspect of the Living Cell (29): physical laws are based on statistical mechanics, which are intimately associated with how systems evolve into disorder.

**Acknowledge**

I am an aerosol scientist mainly for air quality. The cause of initiating this work is I have been told the number size distribution of a system of airborne particles presenting as lognormal distribution since I opened my PhD life in the year of 2006. For long time I was wondering why. Until the year of 2016, I read *information entropy* from somewhere, then started doing this work. Since then, I received a lot of help from my friends, colleagues, and students. Special thanks to my PhD supervisor. Great thanks to all of them for their profound knowledge and enlightening discussions. I am pleasant to share and discuss the ideas with anybody who is interested in this topic.